\documentclass[plaindraft,letter]{jpp-AAS}

\usepackage{graphicx,amsmath}
\usepackage{natbib}
\usepackage[colorlinks,citecolor=blue]{hyperref}
\usepackage{cleveref}

\def\apj{Astrophys.~J.}
\def\apjl{Astrophys.~J.~Lett.}
\def\aap{Astron.~Astrophys.}
\def\araa{Ann.~Rev.~Astron.~Astrophys.}
\def\mnras{Mon.~Not.~R.~Astron.~Soc.}
\def\nat{Nature}

\def\pop{Phys.~Plasmas}
\def\prl{Phys.~Rev.~Lett.}
\def\jgr{J.~Geophys.~Res.}

\def\pop{Phys.~Plasmas}
\def\pof{Phys.~Fluids}

\def\jpp{J.~Plasma Phys.}

\def\ropp{Rev.~Plasma Phys.}
\def\ppcf{Plasma Phys.~Controlled Fusion}
\def\rmp{Rev.~Mod.~Phys.}

\ifCUPmtlplainloaded \else
  \checkfont{eurm10}
  \iffontfound
    \IfFileExists{upmath.sty}
      {\typeout{^^JFound AMS Euler Roman fonts on the system,
                   using the 'upmath' package.^^J}%
       \usepackage{upmath}}
      {\typeout{^^JFound AMS Euler Roman fonts on the system, but you
                   dont seem to have the}%
       \typeout{'upmath' package installed. JPP.cls can take advantage
                 of these fonts, if you use 'upmath' package.^^J}%
       \providecommand\upi{\pi}%
      }
  \else
    \providecommand\upi{\pi}%
  \fi
\fi

\ifCUPmtlplainloaded \else
  \checkfont{msam10}
  \iffontfound
    \IfFileExists{amssymb.sty}
      {\typeout{^^JFound AMS Symbol fonts on the system, using the
                'amssymb' package.^^J}%
       \usepackage{amssymb}%
       \let\le=\leqslant  
       \let\ge=\geqslant  
      }{}
  \fi
\fi

\ifCUPmtlplainloaded \else
  \IfFileExists{amsbsy.sty}
    {\typeout{^^JFound the 'amsbsy' package on the system, using it.^^J}%
     \usepackage{amsbsy}}
    {\providecommand\boldsymbol[1]{\mbox{\boldmath $##1$}}}
\fi

\newcommand{\DD}[2]{\frac{{\rm d}^2 #2}{{\rm d} #1^2}}
\newcommand\bb[1]{\mbox{\boldmath{$#1$}}}

\newcommand{\mc}[1]{\mathcal{#1}}

\newcommand{\mrm}[1]{\mathrm{#1}}

\newcommand{\ex}{\hat{\bb{x}}}
\newcommand{\ey}{\hat{\bb{y}}}
\newcommand{\ez}{\hat{\bb{z}}}

\newcommand{\const}{{\rm const}}
\DeclareMathOperator{\sech}{sech}
\newcommand{\taucs}{\tau_\mrm{cs}}
\newcommand{\taupa}{\tau_\mrm{pa}}
\newcommand{\tauAr}{\tau_\mrm{A,r}}
\newcommand{\Br}{B_\mrm{r}}
\newcommand{\betar}{\beta_\mrm{r}}
\newcommand{\vAr}{v_\mrm{A,r}}
\newcommand{\kt}{k_\mrm{t}}
\newcommand{\ktmax}{k_\mrm{t}^\mrm{max}}
\newcommand{\kmMax}{k_{y\mrm{,m}}^\mrm{max}}
\newcommand{\gammat}{\gamma_\mrm{t}}
\newcommand{\gammatmax}{\gamma_\mrm{t}^\mrm{max}}
\newcommand{\Mach}{M_\mrm{A,0}}
\newcommand{\Nmax}{N_\mrm{max}}

\title[Onset of magnetic reconnection]{Onset of magnetic reconnection in a collisionless, high-beta plasma}

\author[A.~Alt and M.~W.~Kunz]%
{Andrew Alt
and Matthew W.~Kunz\thanks{Email addresses for correspondence: aalt@pppl.gov, mkunz@princeton.edu} }

\affiliation{Department of Astrophysical Sciences, Princeton University, Peyton Hall, Princeton, NJ 08544, USA\\[\affilskip]
Princeton Plasma Physics Laboratory, PO Box 451, Princeton, NJ 08543, USA}

\pubyear{2019}
\volume{}
\pagerange{}
\date{\today}
\begin{document}

\maketitle


\begin{abstract}
In a magnetized, collisionless plasma, the magnetic moment of the constituent particles is an adiabatic invariant. An increase in the magnetic-field strength in such a plasma thus leads to an increase in the thermal pressure perpendicular to the field lines. Above a $\beta$-dependent threshold (where $\beta$ is the ratio of thermal to magnetic pressure), this pressure anisotropy drives the mirror instability, producing strong distortions in the field lines on ion-Larmor scales. The impact of this instability on magnetic reconnection is investigated using a simple analytical model for the formation of a current sheet (CS) and the associated production of pressure anisotropy. The difficulty in maintaining an isotropic, Maxwellian particle distribution during the formation and subsequent thinning of a CS in a collisionless plasma, coupled with the low threshold for the mirror instability in a high-$\beta$ plasma, imply that the geometry of reconnecting magnetic fields can differ radically from the standard Harris-sheet profile often used in simulations of collisionless reconnection. As a result, depending on the rate of CS formation and the initial CS thickness, tearing modes whose growth rates and wavenumbers are boosted by this difference may disrupt the mirror-infested CS before standard tearing modes can develop. A quantitative theory is developed to illustrate this process, which may find application in the tearing-mediated disruption of kinetic magnetorotational ``channel'' modes.
\end{abstract}


%
%
%
\section{Introduction}\label{sec:introduction}

Magnetic reconnection is the process by which magnetic energy is converted to plasma energy via a rapid topological rearrangement of magnetic-field lines \citep{zy09,ykj10,lu16}. It is usually preceded by a slow phase in which magnetic flux is accumulated in an increasingly thin current sheet (CS). Recently, it has been conjectured that this preparatory phase of CS formation, along with the material properties of the host plasma, determine the characteristics of the tearing modes that ultimately disrupt the sheet and thereby set the maximum aspect ratio above which CSs cannot survive \citep{pv14,tenerani15,lu16,ul16,comisso17,huang17}. This maximum aspect ratio is important for (at least) two reasons. First, the large aspect ratio of the Sweet--Parker CS \citep{parker57,sweet58} in high-Lundquist-number plasmas, being violently unstable to the plasmoid instability \citep{loureiro07,bhattacharjee09}, may not be realizable during CS formation. Second, the maximum aspect ratio may define a disruption scale in critically balanced Alfv\'{e}nic turbulence, below which the intense, sheet-like structures become tearing unstable and break up \citep{bl17,lb17a,lb17b,mallet17a,mallet17b}.

All of the work thus far on CS formation and tearing-mediated disruption was either couched within a collisional magnetohydrodynamic (MHD) framework or focused on collisionless plasmas with $\beta\doteq{8}\upi{nT}/B^2\lesssim{1}$ ($n$ is the plasma density, $T$ the temperature, and $B$ the magnetic-field strength). The latter restriction precludes application of those results to many dilute, weakly collisional astrophysical plasmas, whose large temperatures and relatively weak magnetic fields imply $\beta\gg{1}$. For example, $n\sim{10}^{-3}~\mrm{cm}^{-3}$, $T\sim{5}~\mrm{keV}$, and $B\sim{1}~\mu\mrm{G}$ in the hot intracluster medium (ICM) of galaxy clusters imply $\beta\sim{10}^2$ \citep{ct02,sc06}; $n\sim{100}~\mrm{cm}^{-3}$, $T\sim{2}~\mrm{keV}$, and $B\sim{1}~\mrm{mG}$ near the accretion radius of Sgr A$^\ast$ at the Galactic center imply $\beta\sim{10}$ \citep{quataert03,marrone07}. The hallmark of such plasmas is that the embedded magnetic field, while energetically subdominant, nevertheless has a strength tens of orders of magnitude above that required to magnetize the plasma (i.e.~$\Omega_i\tau\ggg{1}$ and $\rho_i\lll{L}$, where $\Omega_i\doteq{eB/m_ic}$ is the ion Larmor frequency, $m_i$ is the ion mass, $\rho_i\doteq{v}_{\mrm{th}i}/\Omega_i$ is the ion Larmor radius, $v_{\mrm{th}i}\doteq(2T/m_i)^{1/2}$ is the ion thermal speed, and $\tau$ and $L$ are representative macroscopic time and length scales, respectively). This hierarchy of scales, particularly in weakly collisional plasmas with collision frequencies $\nu$ satisfying $\nu\tau\ll{1}$, biases the plasma properties with respect to the magnetic-field direction \citep{braginskii65}. Notably, the thermal pressure becomes {\em anisotropic}.

There is a relatively large body of work on the impact of pressure anisotropy on tearing modes \citep{cd81,coppi83,cp84,cl85,ambrosiano86,shi87,karimabadi05,haijima08,quest10,matteini13,gingell15}, as well as on the production and impact of pressure anisotropy during the reconnection process itself \citep{drake06,le09,schoeffler11,egedal13,cassak15,le16}. Here we focus instead on the pressure anisotropy adiabatically produced during the CS formation, prior to the reconnection event. Namely, as the CS thins, the magnetic-field strength in the in-flowing fluid elements increases. An increase in field strength in a weakly collisional, magnetized plasma leads, by adiabatic invariance, to an increase (decrease) in the thermal pressure perpendicular (parallel) to the field lines \citep{cgl56}. Above an $\mc{O}(1/\beta)$ threshold, this pressure anisotropy drives the mirror instability \citep{barnes66,hasegawa69,sk93}, which produces strong distortions in the field lines and traps particles on ion-Larmor scales \citep{kunz14,riquelme15}. In what follows, we ask how the production of pressure anisotropy during CS formation and the consequent triggering of ion-Larmor-scale mirror instabilities in a $\beta\gg{1}$ plasma impacts the onset of tearing-mediated reconnection.

\section{Prerequisites}

%
%
\subsection{CS formation and pressure anisotropy}

We first establish that pressure anisotropy is produced during CS formation. For that, we adopt a simple local model for CS formation based on a one-dimensional generalization of the Chapman--Kendall solution \citep[][\S 2]{ck63,tolman18}. A sheared magnetic field $\bb{B}(x,t)=\Br[x/a(t)]\ey+B_\mrm{g}\ez$ is frozen into an incompressible, time-independent fluid velocity $\bb{u}(x,y)=-(x\ex-y\ey)/2\taucs$, where $\Br$ and $B_\mrm{g}\doteq \theta\Br$ are constants describing the strengths of the reconnecting and guide components of $\bb{B}$, respectively, and $\taucs$ is the characteristic CS-formation timescale. These expressions satisfy the reduced MHD equations provided that the CS half-thickness $a(t)$ and length $L(t)$ satisfy $a(t)/a_0=L_0/L(t)=\exp(-t/\taucs)$, where the ``$0$'' subscript denotes an initial value. This model may be regarded as a Taylor expansion about the neutral line ($x=0$) of a more complicated (e.g.~Harris) CS profile, and so we restrict its validity to $|y|\ll{L(t)}$ and $|x|\lesssim{a(t)}$, beyond which $\bb{B}$ is taken to be spatio-temporally constant. (Indeed, this simple model is only meant to illustrate that $\Delta_p>0$ can be driven during CS formation.) We assume $\sqrt{\rho_{i,\mrm{r}}/a }\ll\theta\lesssim{1}$ and $\Omega_i\taucs\gg{1}$, where $\rho_{i,\mrm{r}}$ is the ion-Larmor radius computed using $\Br$, so that the entire CS is well magnetized (even near $x=0$).\footnote{This guarantees that any particle whose guiding center lies near $x=0$ executes Larmor motion about $B_\mrm{g}$ rather than a betatron orbit with turning points at ${\sim}\sqrt{\rho_{i,\mrm{r}}a}$ \citep[as in][]{dobrowolny68}.}

Using these fields, it is straightforward to show that the magnetic-field strength in a fluid element starting at $x=\xi_0$ (with $|\xi_0|\le{a_0}$) and moving towards $x=0$ is
\begin{equation}
B(\xi(t),t)=\Br\bigl[\theta^2+\exp(t/\taucs)(\xi_0/a_0)^2\bigr]^{1/2} ,
\end{equation}
where $\xi(t)=\xi_0\exp(-t/2\taucs)$ is a Lagrangian coordinate co-moving with the fluid element. This change in $B$ drives field-aligned pressure anisotropy, $\Delta_p\doteq{p}_\perp/p_\parallel-1$, adiabatically in the fluid frame. Using $\mu$ conservation in the form $p_\perp\propto{B}$ and assuming $\Delta_p(x,t=0)=0$,
\begin{equation}
\Delta_p(\xi(t),t)	= \biggl[\frac{\theta^2+\exp(t/\taucs)(\xi_0/a_0)^2}{\theta^2+(\xi_0/a_0)^2}\biggr]^{1/2}-1 \approx \frac{t}{2\taucs} \frac{(\xi_0/a_0)^2}{\theta^2+(\xi_0/a_0)^2}\doteq\frac{t}{\taupa} 
\label{eqn:Deltaapprox}
\end{equation}
for $t/\taucs\ll{1}$.\footnote{If the second adiabatic invariant, $J$, were also conserved -- unlikely in a $\beta\gg{1}$ plasma with Alfv\'{e}nic, incompressible flows -- the exponent $1/2$ in \eqref{eqn:Deltaapprox}  becomes $3/2$ and $\taupa$ changes by an inconsequential factor of $3$.} Thus, pressure anisotropy increases in all fluid elements.

If nothing interferes with the adiabatic increase in pressure anisotropy, the plasma in a fluid element will eventually become mirror unstable when $\Delta_p\gtrsim{1}/\beta_\perp$, where
\begin{equation}
\beta_\perp(\xi(t),t) = \beta_0(\xi_0)\biggl[\frac{\theta^2+(\xi_0/a_0)^2}{\theta^2+\exp(t/\taucs)(\xi_0/a_0)^2}\biggr]^{1/2} \approx \beta_0 \biggl(1-\frac{t}{3\taupa}\biggr)
\label{eqn:betaapprox}
\end{equation}
is the adiabatically evolving perpendicular plasma beta in the fluid frame ($\beta_0$ is its initial value). Comparing (\ref{eqn:Deltaapprox}) and (\ref{eqn:betaapprox}), this occurs at $t_\mrm{m}\sim\taupa/\beta_0$ for $\beta_0\gg{1}$. If the guide field is small compared to the local reconnecting field ($\theta\ll\xi_0/a_0$), this time is a small fraction of the CS-formation time scale, $t_\mrm{m}\sim\taucs/\beta_0$, and so the CS becomes mirror-unstable early in its evolution. With a larger guide field ($\theta\gg\xi_0/a_0$), $t_\mrm{m}\sim\taucs(a^2_0/\xi^2_0)(\theta^2/\beta_0)$. This time is also early in the CS evolution for $\xi_0\lesssim{a}_0$, since $\theta\ll\beta^{1/2}$ is required in this model for the plasma to reliably exceed the mirror-instability threshold.\footnote{\label{fn:thetaLim} If the asymptotic value of the reconnecting field, $\Br$, is constant, then the maximum change of $B$ in a fluid element is bounded, $B(t)/B(0)< (1+\theta^{-2})^{1/2}$, and so $\Delta_p<(1+\theta^{-2})^{1/2}-1$. Therefore, $\theta\lesssim\beta^{1/2}$ is required to reach the mirror threshold. In other models where $\Br$ increases in time \citep[e.g.][]{tolman18}, no such limit on $\theta$ exists.}

These times must be compared to the characteristic time scales for tearing modes that facilitate magnetic reconnection in the forming CS. Before doing so, we review the basic properties of the mirror instability.

%
%
\subsection{Mirror instability}

As $B$ increases, adiabatic invariance drives $\Delta_p>0$, with plasma becoming mirror-unstable when $\Lambda_\mrm{m}\doteq\Delta_p-1/\beta_\perp>0$. Just beyond this threshold ($0<\Lambda_{\rm m}\ll{1}$), oblique modes with wavenumbers $k_{\parallel,\mrm{m}}\rho_i\sim (k_{\perp,\mrm{m}}\rho_i)^2\sim\Lambda_\mrm{m}$ and polarization $\delta{B}_\perp/\delta{B}_\parallel\sim\Lambda_\mrm{m}^{1/2}$ grow exponentially at a maximum rate $\gamma_\mrm{m}\sim\Omega_i\Lambda_\mrm{m}^2$ \citep{hellinger07}. Once this growth rate becomes larger than the rate at which $\Delta_p$ is produced ($\gamma_\mrm{m}\taupa\gtrsim{1}$), the growth of $\Delta_p$ stops. This yields a maximum mirror-instability parameter,  $\Lambda_\mrm{m}\gtrsim(\Omega_i\taupa)^{-1/2}\doteq\Lambda_\mrm{m,max}$. Kinetic simulations show that, once $\Lambda_\mrm{m}(t)\sim\Lambda_\mrm{m,max}$, mirrors rapidly drain $\Lambda_\mrm{m}(t)\rightarrow{0}^+$ and attain amplitudes $\delta{B}_\parallel/B\sim\Lambda^{1/2}_\mrm{m,max}$ \citep{kunz14}. This is the end of the linear stage; for $\beta_0\gg{1}$, this occurs at $t/\taupa\sim{1}/\beta_0+\Lambda_\mrm{m,max}$.

As the CS continues to thin, $\Delta_p>0$ is continuously driven. Mirror modes then maintain marginal stability ($\Lambda_\mrm{m}\simeq{0}^+$) by growing secularly, $\delta{B}^2_\parallel\propto{t}^{4/3}$, and trapping an increasing fraction of particles \citep{schekochihin08,kunz14,rincon15}. Independent of $\Lambda_\mrm{m,max}$, saturation occurs at $t\sim\taupa$ and $\delta{B}/B\sim{1}$, when these particles pitch-angle scatter off sharp bends in the magnetic field occurring at the mirror boundaries at a rate $\nu_\mrm{m}\sim\beta/\taupa$; this maintains marginal stability by severing the adiabatic link between $\Delta_p$ and changes in $B$ \citep{kunz14,riquelme15}. Thereafter, $\Delta_p\simeq{1}/\beta_\perp$, even as $B$ changes.

This evolution was found for situations in which $\taupa$ is comparable to the dynamical time in the system (e.g., linear shear flows). However, for locations $\xi_0\ll\theta{a}_0$ deep inside the CS, $\taupa\gg\taucs$. In this case, local mirror growth cannot outpace CS formation, and any potential mirrors are advected and distorted faster than they can grow. When $\theta\gg{1}$, $\taupa\gg\taucs$ in the entire CS. We thus focus only on cases with $\theta\lesssim{1}$ and locations $\xi_0\gtrsim\theta{a}_0$.

%
%
\subsection{Collisionless tearing instability} 

Next we review the theory of collisionless tearing modes, applicable when the inner-layer thickness of the tearing CS, $\delta_\mrm{in}\lesssim\rho_e$. To determine under what condition this criterion is satisfied, we use standard MHD tearing theory \citep{fkr63} to estimate
\begin{equation}\label{eqn:din}
\delta_\mrm{in}^\mrm{MHD}=\bigl[\gammat(\kt\vAr)^{-2}a^2\eta\bigr]^{1/4}=a\bigl[\gammat\tauAr(\kt{a})^{-2}S^{-1}_a\bigr]^{1/4},
\end{equation}
where $\vAr\doteq\Br/(4\upi{m}_in_i)^{1/2}$ is the Alfv\'{e}n speed of the reconnecting field, $\tauAr\doteq{a}/\vAr$ is the Alfv\'{e}n crossing time of the CS, $\eta$ is the (collisional) resistivity, and $S_a\doteq{a}\vAr/\eta$ is the Lundquist number. Using an estimate for the growth rate $\gammat$ of the fastest-growing collisional tearing mode with wavenumber $\kt$ oriented along the CS \citep{fkr63,coppi76,ul16}, the validity condition for collisionless tearing theory to hold becomes 
\begin{equation}\label{eqn:Scollisionless}
S_a\gtrsim(a/\rho_e)^4.
\end{equation}
This gives $a\lesssim{10}^{-6}~\mrm{pc}$ for the ICM parameters listed in \S\ref{sec:introduction}, a satisfiable constraint given that $\rho_i\sim{10}^{-9}~\mrm{pc}$ and the outer scale of ICM magnetic-field fluctuations is observationally inferred to be ${\sim}10~\mrm{kpc}$ \citep{ev06,guidetti08,bonafede10,vacca12,govoni17}, comparable to the collisional mean free path. At the accretion radius of Sgr A$^\ast$, this constraint is $a\lesssim{10}^{-10}~\mrm{pc}$, which is ${\sim}10^2$ larger than $\rho_i$ and ${\sim}10^8$ times smaller than the collisional mean free path. As long as \eqref{eqn:Scollisionless} is satisfied (which becomes easier as $a$ shrinks), $\gammat$ and $\kt$ are estimated as follows.

In a $\beta\gtrsim{1}$ plasma when the tearing-mode instability parameter $\Delta'(\kt)$ \citep{fkr63} is small, satisfying $\Delta' \delta_\mrm{in} \sim (\Delta' d_e)^2 \ll 1$ (``FKR-like''; \citet{karimabadi05}),
\begin{equation}\label{eqn:gammat_smallDelta}
\gammat^{\rm FKR}\tauAr\sim\biggl(\frac{m_e}{m_i}\biggr)^{1/2}\biggl(\frac{d_i}{a}\biggr)^2 \,\kt{a}\,\Delta'a ,
\end{equation}
where $d_e$ and $d_i\doteq\rho_i/\beta^{1/2}_{i}=d_e(m_i/m_e)^{1/2}$ are, respectively, the electron and ion skin depths \citep{fitzpatrick04,fitzpatrick07}. (Our CS formation model leaves $d_e,d_i$ constant.) This growth rate is approximately independent of $\kt$ in a Harris sheet, for which $\Delta'a=2(1/\kt{a}-\kt{a})\sim(\kt{a})^{-1}$ at $\kt{a}\ll{1}$. The large-$\Delta'$ (``Coppi-like'') growth rate satisfies
\begin{equation}\label{eqn:gammat_largeDelta}
\gammat^{\rm Coppi}\tauAr\sim\biggl(\frac{m_e}{m_i}\biggr)^{1/5}\biggl(\frac{d_i}{a}\biggr)\,\kt{a},
\end{equation}
independent of $\Delta'$ \citep{fitzpatrick07}. An estimate for $\gammat$ and $\kt$ of the fastest-growing Coppi-like mode in a Harris sheet can be obtained by balancing \eqref{eqn:gammat_smallDelta} and \eqref{eqn:gammat_largeDelta}:
\begin{subequations}\label{eqn:tearingn}
\begin{align}
\gammatmax\tau_{\rm A,r}	&\sim \biggl(\frac{m_e}{m_i}\biggr)^{1/2}\biggl(\frac{d_i}{a}\biggr)^{2},\label{eqn:tearingn:g}\\*
\ktmax a 					&\sim \biggl(\frac{m_e}{m_i}\biggr)^{3/10}\biggl(\frac{d_i}{a}\biggr) .\label{eqn:tearingn:k}
\end{align}
\end{subequations}
These modes are the fastest growing provided they fit into the length $L$ of the CS, i.e.~$\ktmax L>1$. Otherwise, the fastest-growing mode is FKR-like.\footnote{\citet{fitzpatrick04,fitzpatrick07} obtained \eqref{eqn:gammat_smallDelta} and \eqref{eqn:gammat_largeDelta} using a two-fluid model assuming cold ions and that the compressional Alfv\'{e}n wave propagates much faster than any other wave in the system (as it would in a high-$\beta$ plasma), thus guaranteeing pressure balance along field lines and nearly incompressible flow. The former (small-$\Delta'$) growth rate agrees with the corresponding kinetic expression in \citet[][their equation (16)]{dl77a} up to a factor of $1/\sqrt{1+\beta_\mrm{g}}$, which is ${\sim}1$ given those authors' assumption of small $\beta$ and large guide field. Both results assumed a Maxwellian background. Alternatively, \citet{cp84} allowed for a spatially uniform $\Delta_p\ne{0}$ in their linear kinetic tearing calculation, but assumed $B_\mrm{g}=0$ and thus obtained different scalings after accounting for axis-crossing particle orbits (see also \citet{cl85} and \citet{quest10}). While we have opted to use the \citet{fitzpatrick04,fitzpatrick07} expressions for $\gamma_{\rm t}$, our analysis can be generalized for any alternative scalings without a significant change in the main qualitative conclusions summarized in \S\ref{sec:summary}. The ``FKR-like'' and ``Coppi-like'' designations are adaptations of those introduced by \citet{ul16}.}

In what follows, we {\em assume} that pressure anisotropy does not appreciably modify these growth rates. This is because saturated mirrors maintain $\Delta_p\simeq{1}/\beta_\perp\ll{1}$, and so the resulting viscous stress effectively enhances the magnetic tension responsible for driving the tearing by a factor of only ${\simeq}3/2$. Other works that postulate an initial $\Delta_p$ (customarily taken to be uniform and thus non-zero even at $x=0$) do not consider its rapid regulation by the mirror instability prior to the onset of tearing, and the enhanced $\gammat$ often found in linear calculations when $\Delta_p > 0$ is largely because the assumption $B_\mrm{g}=0$ permits axis-crossing particle orbits in the inner regions of the CS and allows threshold-less instabilities such as the Weibel instability \citep[e.g.][]{cp84}.

%
%
\section{Reconnection onset when $\Delta_p=0$} 

Before determining how mirror-unstable pressure anisotropy affects a gradually forming CS, we recapitulate the theory of CS disruption by tearing modes \citep{pv14,tenerani15,lu16,ul16}, specialized to the case of collisionless tearing in a high-$\beta$ plasma. That is, we ignore the production of pressure anisotropy during CS formation and instead determine when $L/a$ has increased enough for tearing modes to prompt reconnection.

\begin{figure}
\centering
\includegraphics[width=0.75\textwidth]{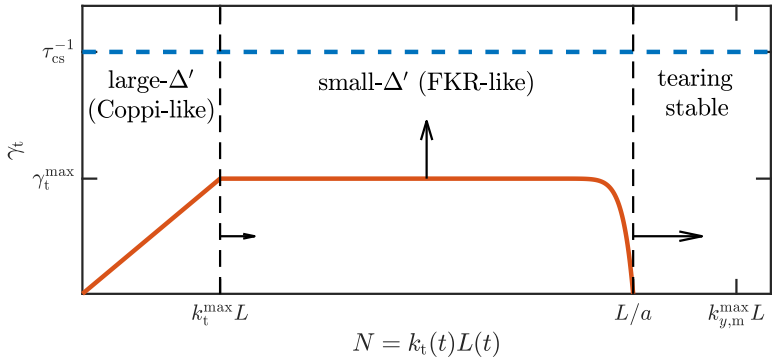}
\caption{Qualitative plot of tearing growth rate $\gammat$ vs.~mode number $N$ (see \eqref{eqn:gammat_smallDelta} and \eqref{eqn:gammat_largeDelta}) shortly after mirror production at $k^\mrm{max}_{y,\mrm{m}}a>{1}$. Arrows indicate evolution as the CS aspect ratio ($L/a$) increases, with $\gammat$ approaching $\taucs^{-1}$ (blue dashed line), $\kt$ approaching the large-$\Delta'$ regime ($\kt\lesssim\ktmax$), and mirrors affecting an increasing number of tearing modes (those with $\kt\gtrsim{k}^\mrm{max}_{y,\mrm{m}}$).}
\label{fig:gamma-t0}
\end{figure}

As the CS's aspect ratio $L/a$ increases in time, modes with progressively larger mode number $N\doteq\kt(t)L(t)=\const$ become unstable and undergo linear evolution with $\gammat(N,t)$ increasing (see figure \ref{fig:gamma-t0}). \citet{ul16} argued that the first tearing mode $N$ to reach the end of its linear stage at the critical time $t_\mrm{cr}(N)$ (when $\gammat\taucs\gtrsim{1}$, neglecting logarithmic corrections \citep{comisso17}) will also be the first to undergo X-point collapse (defined by when the island width $w\sim{1}/\Delta'$) and, soon thereafter, disrupt the CS ($w\sim{a}$). We adopt this argument and estimate the CS disruption time $t_\mrm{disrupt}$ for a collisionless Harris sheet with $L(t)a(t)=\const$. (The same procedure can be used to investigate alternative CS profiles and evolution.) Note that, for the Harris-sheet profile, $\gammat^{\rm FKR}\approx\gammatmax$ for $\kt{a}\ll{1}$ (see \eqref{eqn:gammat_smallDelta} and \eqref{eqn:tearingn:g}), so the only difference between these modes are their wavenumbers and, thus, their $\Delta'\sim{1}/\kt{a}^2$.

%
%

Each unstable mode $N$ starts in the small-$\Delta'$ (``FKR-like") regime ($N>\Nmax(t)$), with $\gammat$ roughly independent of $\kt$ for $\kt{a}\ll{1}$. However, because $\Nmax\propto(L/a)(d_i/a)\propto{a}^{-3}$ increases in time, these FKR-like modes approach the large-$\Delta'$ (``Coppi-like'') regime, making the transition at $t=t_\mrm{tr}(N)$ when
\begin{equation}
\frac{a(t_\mrm{tr}(N))}{a_0}\sim\biggl(\frac{m_e}{m_i}\biggr)^{1/10}\biggl(\frac{L_0d_i}{a^2_0}\biggr)^{1/3}N^{-1/3}.
\end{equation}
Larger $N$ corresponds to larger $t_\mrm{tr}(N)$, and so the first mode to make this transition is $N=1$; i.e.~at $t=t_\mrm{tr}(1)$, the fastest Coppi-like mode (see \eqref{eqn:tearingn:k}) just fits inside the CS. All modes satisfying $\ktmax a\lesssim\kt{a}\ll{1}$ obtain growth rates $\gammat\taucs\gtrsim{1}$ at roughly the same time, $t=t_\mrm{cr}$, when (using \eqref{eqn:tearingn:g})
\begin{equation}\label{eqn:tearingDisrupts}
\frac{a(t_\mrm{cr})}{a_0}\lesssim\biggl(\frac{m_e}{m_i}\biggr)^{1/6}\biggl(\frac{d_i}{a_0}\biggr)^{2/3}\Mach^{-1/3} ,
\end{equation}
where $\Mach\doteq\tauAr(t=0)/\taucs$ is the initial Alfv\'{e}nic Mach number of the CS formation. These modes have
\begin{equation}\label{eqn:tearingDisrupts_N}
\frac{L(t_\mrm{cr})}{a(t_\mrm{cr})}\gg{N}\ge{N}_\mrm{cr}\doteq\biggl(\frac{m_e}{m_i}\biggr)^{-1/5}\biggl(\frac{L_0}{d_i}\biggr)\Mach.
\end{equation}
This is an important distinction from the collisional MHD case, in which larger $N>N_\mrm{cr}$ corresponds to larger $t_\mrm{cr}(N)$ (since $\gammat^{\rm FKR}\propto\kt^{-2/5}$ at $\kt{a}\ll{1}$ instead of $\kt^{0}$).

Another important distinction from the MHD case lies in the nonlinear evolution, during which the MHD FKR modes behave differently than the MHD Coppi modes. While the latter are expected to rapidly evolve towards X-point collapse soon after $t=t_\mrm{cr}$ due to their large $\Delta'$, the former undergo secular ``Rutherford'' evolution that increases $\Delta'(k_N)w_N$ for a given mode $N$ until $w_N\sim{1}/\Delta'$ \citep{rutherford73,waelbroeck89,waelbroeck93,loureiro05,arcis09}. However, in the collisionless case, the FKR-like modes reach $\gammat\taucs\sim{1}$ at the same time as the fastest Coppi-like mode. If the latter is accessible, then the fastest-growing mode $\Nmax$ already has $\Delta'd_e\sim{1}$ at $t_\mrm{cr}(\Nmax)$ and so X-point collapse likely occurs soon after \eqref{eqn:tearingDisrupts} is satisfied. The CS is then said to be ``disrupted'' at $t_\mrm{disrupt}\sim{t}_\mrm{cr}(\Nmax)$. For there to be no Coppi-like modes when \eqref{eqn:tearingDisrupts} is satisfied (i.e.~$N_\mrm{cr}<{1}$), $\Mach\lesssim(m_e/m_i)^{1/5}(d_i/L_0)$, a rather stringent condition that is difficult to satisfy when $\beta_0\gg{1}$ and $\rho_{i0}/L_0\ll{1}$. 

That being said, given the uncertainties in the nonlinear evolution of collisionless tearing modes in a high-$\beta$, magnetized plasma -- especially regarding the existence (or nonexistence) of a secular ``Rutherford'' phase and the production of pressure anisotropy during X-point collapse -- we focus primarily on the critical time for reconnection onset (when $\gammat\taucs\gtrsim{1}$) rather than the CS disruption time (when $w\sim{a}$).\footnote{Another reason for prudence is Drake \& Lee's (1977{\it b})\nocite{dl77b} argument that single-mode tearing with a guide field saturates via trapped-electron effects with an amplitude comparable to the inner-layer thickness, $w\sim\delta_\mrm{in}$. This argument was confirmed, and refined by incorporating finite-Larmor-radius effects, by \citet{karimabadi05}.}

%
%
\begin{figure}
\centering
\includegraphics[width=0.6\textwidth]{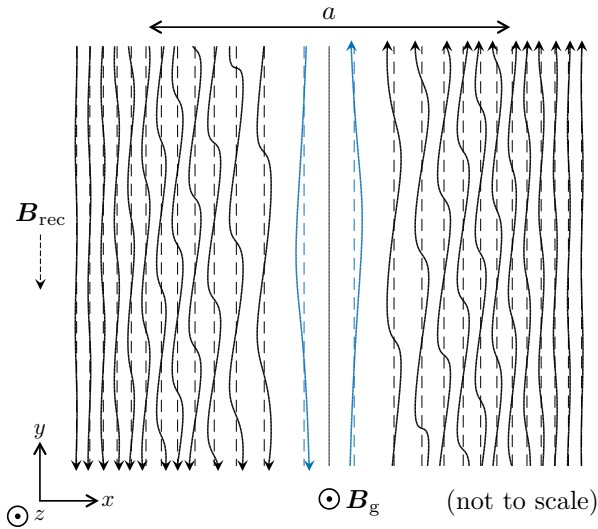}
\caption{Qualitative illustration of magnetic-field lines in an evolving, mirror-infested Harris CS with $\theta \ll 1$.}
\label{fig:CS_with_mirrors}
\end{figure}
%
%
%

%
%
\section{Reconnection onset when $\Delta_p \ne 0$}

We now consider the effects of  mirrors on an evolving CS subject to tearing modes. Because different portions of the CS have different $\rho_i$ and $\taupa$, there will be a range of mirror wavenumbers, $k_{y,\mrm{m}}(x)$, along the CS (see figure \ref{fig:CS_with_mirrors}).

The smallest $k_{y,\mrm{m}}$ will be located the nearest to $x=0$ where mirrors can form, since these regions have the largest values of $\rho_i$ and $\taupa$. We argue that, since tearing modes with wavenumbers $\kt$ much smaller than this $k_{y,\mrm{m}}^\mrm{min}$ will see a rapidly $y$-varying magnetic field that averages to its unperturbed value, these modes are likely unaffected by the mirrors (or at least less affected than other modes). The largest $k_{y,\mrm{m}}$ will be located near $|x|\sim{a}$, where $\rho_i$ and $\taupa$ are at their smallest values. All tearing modes with $\kt\gg\kmMax$ will see an approximately uniform-in-$y$ magnetic field, but will have their $\Delta'(\kt)$ enhanced by the mirrors' effect on the $x$-variation of the CS profile. If the CS is able to stretch to the point where $\kmMax\lesssim\ktmax$ before the onset of tearing, then all of the modes that are unaffected by the mirrors will have smaller growth rates and thus be unimportant for CS reconnection. The condition $\kmMax\lesssim\ktmax$ is thus a sufficient (but not necessary) condition for mirrors to matter.

We now follow the evolution of $k_{y,\mrm{m}}^\mrm{max}$ as the CS evolves, and investigate the evolution of tearing modes with $\kt\gg{k}_{y,\mrm{m}}^\mrm{max}$. We treat two cases, depending upon the size of the guide field and thus the component of the mirrors' wavevector along the CS at $|x|\sim{a}$,
\begin{equation}\label{eqn:ky}
k_{y,\mrm{m}}\sim k_{\parallel,\mrm{m}}\frac{B_{\rm r}}{B}+k_{\perp,\mrm{m}}\frac{B_{\rm g}}{B}=k_{\parallel,\mrm{m}} \frac{B_{\rm r}}{B}\biggl(1+\theta\frac{k_{\perp,\mrm{m}}}{k_{\parallel,\mrm{m}}}\biggr) .
\end{equation}
With $k_{\perp,\mrm{m}}/k_{\parallel,\mrm{m}}\sim\Lambda^{-1/2}_\mrm{m,max}$ for the fastest-growing mirror mode, we have $k_{y,\mrm{m}}\sim{k}_{\parallel,\mrm{m}}$ for $\theta\ll\Lambda^{1/2}_\mrm{m,max}$ and $k_{y,\mrm{m}}\sim\theta{k}_{\perp,\mrm{m}}$ for $\Lambda^{1/2}_\mrm{m,max}\ll\theta\lesssim{1}$. (In both cases, $\Lambda_\mrm{m,max}\sim(d_i/a_0)^{1/2}\Mach^{1/2}$.)

%
%
\subsection{When mirrors affect tearing if $\theta\ll\Lambda^{1/2}_\mrm{m,max}$}\label{sec:smalltheta}

At $x\sim{a}$, the local reconnecting field is near its asymptotic value and $\taupa\sim\taucs$. Starting at time $t_\mrm{m}\sim\taucs/\beta_0\ll\taucs$, unstable mirror modes grow rapidly at this location ($a$ and $\tauAr$ hardly change from their initial values in a time $t_\mrm{m}$.) Unless tearing modes disrupt the CS within $t_\mrm{disrupt}\lesssim\taucs$ -- which is extremely unlikely, requiring \eqref{eqn:tearingDisrupts} to be satisfied within $\taucs$ -- these mirrors will saturate with $\delta{B}\sim\Br$ and 
\begin{equation}\label{eqn:kpar}
k_{y,\mrm{m}}^\mrm{max}(t)\rho_i\sim\frac{L_0}{L(t)}(\Omega_i\taucs)^{-1/2}\sim\frac{a(t)}{a_0}\biggl(\frac{d_i}{a_0}\biggr)^{1/2}\Mach^{1/2},
\end{equation}
where we have accounted for the Lagrangian stretching of the perturbations during CS formation.

To determine the effect of these mirrors on tearing, it is useful (as argued above) to first establish when $k^\mrm{max}_{y,\mrm{m}}(t)$ enters the large-$\Delta'$ regime in which $\gammat\propto{k}$ (the leftmost portion of figure \ref{fig:gamma-t0}), i.e.~when the mirrors influence the fastest-growing tearing modes. Combining \eqref{eqn:tearingn:k} and \eqref{eqn:kpar}, we find that $a(t)$ must satisfy
\begin{equation}\label{eqn:mirrorsCoppi1}
\frac{a(t)}{d_i}\lesssim\biggl(\frac{m_e}{m_i}\biggr)^{1/10}\biggl(\frac{d_i}{a_0}\biggr)^{-1/2}\beta_0^{1/6}\Mach^{-1/6}
\end{equation}
for $k^\mrm{max}_{y,\mrm{m}}(t)\lesssim\ktmax(t)$. Equation (\ref{eqn:mirrorsCoppi1}) happens before the sheet would be disrupted in the absence of mirrors (see (\ref{eqn:tearingDisrupts})) if
\begin{equation}\label{eqn:CoppiWins1}
\frac{a_0}{d_i}\gtrsim\biggl(\frac{m_e}{m_i}\biggr)^{2/5}\beta_0^{-1}\Mach^{-1},
\end{equation}
which is easily satisfied under the conditions of interest. Thus, there will be a time at which all tearing modes with $\kt\gtrsim\ktmax$ are affected by mirrors. How the tearing progresses after \eqref{eqn:mirrorsCoppi1} is satisfied will be discussed once the corresponding conditions for the other $\theta$-regime are derived.

%
%
\subsection{When mirrors affect tearing if $\Lambda^{1/2}_\mrm{m,max}\ll\theta\lesssim{1}$} 

As $B_\mrm{g}$ is increased, things will continue in much the same way as in \S\ref{sec:smalltheta} except that the initial $k^\mrm{max}_{y,\mrm{m}}\sim\theta{k}_{\perp,\mrm{m}}$. That is, equation \eqref{eqn:kpar} is replaced by
\begin{equation}\label{eqn:kperp}
k_{y,\mrm{m}}^\mrm{max}(t)\rho_i\sim\frac{L_0}{L(t)}\theta(\Omega_i\taucs)^{-1/4}\sim\frac{a(t)}{a_0}\biggl(\frac{d_i}{a_0}\biggr)^{1/4}\theta\Mach^{1/4}.
\end{equation}
This means that the condition on $a(t)$ that $k^\mrm{max}_{y,\mrm{m}}(t)\lesssim\ktmax(t)$ (cf.~\eqref{eqn:mirrorsCoppi1}) becomes 
\begin{equation}\label{eqn:mirrorsCoppi2}
\frac{a(t)}{d_i}\lesssim\biggl(\frac{m_e}{m_i}\biggr)^{1/10}\biggl(\frac{d_i}{a_0}\biggr)^{-5/12}\theta^{-1/3}\beta_0^{1/6}\Mach^{-1/12}.
\end{equation}
If the initial state satisfies
\begin{equation}\label{eqn:CoppiWins2}
\frac{a_0}{d_i}\gtrsim\biggl(\frac{m_e}{m_i}\biggr)^{4/5}\theta^4\beta_0^{-2}\Mach^{-3},
\end{equation}
then \eqref{eqn:mirrorsCoppi2} occurs before \eqref{eqn:tearingDisrupts}, when the sheet would be disrupted without the mirrors.

%
%
\subsection{Mirror-stimulated onset of reconnection} 

If either \eqref{eqn:CoppiWins1} or \eqref{eqn:CoppiWins2} is satisfied, then mirrors influence all tearing modes before they could otherwise disrupt the CS in the absence of mirrors. We now quantify that influence, focusing on those tearing modes with $\kt\gg{k}^\mrm{max}_{y,\mrm{m}}$ (see \eqref{eqn:kpar} and \eqref{eqn:kperp}). As argued previously, these modes see a magnetic field that is roughly uniform in $y$ but is rapidly varying in $x$ due to the mirrors, with an initial $k_{x,\mrm{m}}\sim{k}_{\perp,\mrm{m}}$ that is then compressed by the CS formation with $k_{x,\mrm{m}}(t)a(t)\sim\const$. This rapid variation enhances $\gammat(\kt)$ for these modes due to the smaller effective sheet thickness (estimated below), which affects both $\Delta'(\kt)$ and the Alfv\'{e}n-crossing time $\tauAr$ (see \eqref{eqn:gammat_smallDelta} and \eqref{eqn:gammat_largeDelta}). 

\subsubsection{Model for a mirror-infested CS}

We argue that $\tauAr$ changes by a small amount, since mirrors modify $\mrm{d}B_y/\mrm{d}x|_{x=0}$ by only a factor of order unity. To determine how $\Delta'(k)$ is modified, we adopt the following simple model for the magnetic-field profile of a mirror-infested Harris CS:
\begin{equation}\label{eqn:Bymodel}
B_y(x)=\Br\tanh\Bigl(\frac{x}{a}\Bigr)\Bigl[1+\varepsilon\sin\Bigl(2k_\mrm{max}a\sech\Bigl(\frac{x}{a}\Bigr)\Bigr)\Bigr] ,
\end{equation}
where $k_\mrm{max}\gg{a}^{-1}$ is a parameter characterizing the peak $k_{x,\mrm{m}}$ occurring at the edge of the CS. This is a WKB approximation describing saturated mirrors with amplitude $\varepsilon\sim\mathcal{O}(1)$ times the local reconnecting field and wavenumber in the $x$-direction given by $k_x(x)=2k_\mrm{max}\sech(x/a)\tanh(x/a)$. This model was chosen because $k_x(x=0)=0$, $k_x(x\to\infty)\to{0}$, and $k_x(x)$ is maximal near the edge of the CS, as anticipated. (What follows is not particularly sensitive to this choice of $k_x(x)$.)

\begin{figure}
\centering
\includegraphics[width=0.65\textwidth]{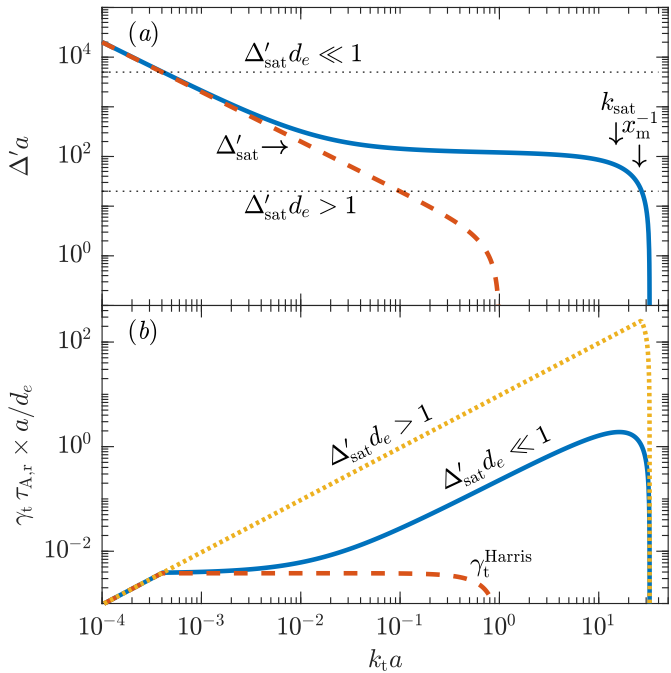}%
\caption{({\it a}) $\Delta'(\kt)$ and ({\it b}) $\gammat(\kt)$ for a Harris CS (red dashed line) and its mirror-infested counterpart (blue solid line), using $k_\mrm{max}a=200\upi$ and $\varepsilon=1/2$ in Eq.~\eqref{eqn:Bymodel}. $\Delta'$ rises rapidly at $\kt x_\mrm{m}\lesssim{1}$ and plateaus for ${k}_\mrm{sat}\gtrsim\kt\gtrsim{1}/(\Delta'_\mrm{sat}a^2)$. Mirror-stimulated tearing thus peaks at $\kt\sim{k}_\mrm{sat}$, regardless of whether $\Delta'_\mrm{sat}d_e\ll{1}$ (blue solid line) or $\Delta'_\mrm{sat}d_e\gtrsim1$ (orange dotted line).}
\label{fig:Deltap-vs-k}
\end{figure}

The resulting $\Delta'(\kt)$ is obtained by numerically integrating the outer differential equation for the flux function, $\psi$ \citep{fkr63}:
\begin{equation}\label{eqn:psi}
\DD{x}{\psi}-\biggl(k^2+\frac{B_y''}{B_y}\biggr)\psi=0,
\end{equation}
with $B_y(x)$ given by \eqref{eqn:Bymodel}. Then $\Delta'\doteq\mrm{d}\ln\psi/\mrm{d}x|_{x=0}$ for the solution that obeys reasonable boundary conditions; an example result is shown in figure \ref{fig:Deltap-vs-k}({\it a}). (Its shape does not change significantly as $\varepsilon$ and $k_\mrm{max}$ vary.) Generally, $\Delta'>0$ for $\kt$ smaller than the inverse of the effective sheet thickness, $a_\mrm{eff}$, which we identify with the location $x_\mrm{m}$ of the peak in $B_y(x)$ closest to $x=0$ (i.e.~the location of the innermost mirror). As $\kt$ decreases from this value, $\Delta'(\kt)$ rises sharply to saturate at $\kt=k_\mrm{sat}$ with value $\Delta'_\mrm{sat} \sim 1/a_\mrm{eff} \sim 1/x_\mrm{m}$, at which it is approximately constant until it nears the Harris-sheet $\Delta'(\kt)\sim{1}/\kt{a}^2$, which it then follows.

The corresponding $\gamma_\mrm{t}(\kt)$ shown in figure \ref{fig:Deltap-vs-k}({\it b}) depends on whether or not $\Delta'_\mrm{sat}d_e\ll{1}$. However, the maximum growth rate always occurs at $k_\mrm{sat}\sim{1}/x_\mrm{m}$, because of the $\kt$-dependence of \eqref{eqn:gammat_smallDelta} and \eqref{eqn:gammat_largeDelta}. Thus, to determine the new $t_\mrm{cr}$, we must calculate $x_\mrm{m}$. This yields two cases based on the size of $\theta$.

%
%

%
%
\subsubsection{Mirror-stimulated tearing for $\theta\ll{x}_\mrm{m}/a$}

When the reconnecting field is the dominant field on the scale of the innermost mirrors, the total ion-cyclotron frequency is $\Omega_i\sim(x_\mrm{m}/a)\Omega_{i,\mrm{r}}$ and $\taupa\sim\taucs$. The $x$-wavenumber of the mirrors at that location is then
\begin{equation}\label{eqn:kxm}
k_{x,\mrm{m}}(t,x_\mrm{m}(t))\rho_{i,\mrm{r}}\sim\biggl(\frac{x_\mrm{m}}{a(t)}\biggr)^{3/4}\biggl(\frac{d_i}{a_0}\biggr)^{1/4}\frac{a_0}{a(t)}\Mach^{1/4},
\end{equation}
where we have accounted for the Lagrangian compression due to CS formation. The innermost mirror is located at $x_\mrm{m}\sim{k}^{-1}_{x,\mrm{m}}$, an $x$-wavelength away from the center. Substituting this into \eqref{eqn:kxm} yields
\begin{equation}\label{eqn:xm}
\frac{x_\mrm{m}}{a(t)}\sim\biggl(\frac{d_i}{a_0}\biggr)^{3/7}\beta_\mrm{r}^{2/7}\Mach^{-1/7}.
\end{equation} 
For this estimate to be self-consistent, we require $\theta\ll{x}_{\rm m}/a$ or, using \eqref{eqn:xm},
\begin{equation}\label{eqn:smallThReq}
\theta\ll\biggl(\frac{d_i}{a_0}\biggr)^{3/7}\beta_\mrm{r}^{2/7}\Mach^{-1/7}.
\end{equation}
Provided this is satisfied, the fastest-growing tearing mode, having $\gammat(k_{\rm sat})$, is either FKR-like, if $d_e/x_\mrm{m}\ll{1}$, or Coppi-like, if $d_e/x_\mrm{m}\gtrsim{1}$. 

In the former case, the maximum tearing growth rate is (using \eqref{eqn:gammat_smallDelta} with $\kt\sim{1}/x_\mrm{m}$ and $\Delta'\sim{1}/x_\mrm{m}$)
\begin{equation}\label{eqn:gammaFKRsmallTh}
\gamma_\mrm{t,m}^\mrm{FKR}\tauAr\sim\biggl(\frac{m_e}{m_i}\biggr)^{1/2}\Biggl(\frac{d_i^{8/7}a_0^{6/7}}{a^2}\Biggr)\betar^{-4/7}\Mach^{2/7} .
\end{equation}
The critical time for onset, $t^\mrm{FKR}_\mrm{cr}$, occurs when $\gamma_\mrm{t,m}^\mrm{FKR}\taucs\sim{1}$, or
\begin{equation}\label{eqn:FKRdisrupts}
\frac{a(t^\mrm{FKR}_\mrm{cr})}{a_0}\lesssim\biggl(\frac{m_e}{m_i}\biggr)^{1/6}\biggl(\frac{d_i}{a_0}\biggr)^{8/21}\betar^{-4/7}\Mach^{-5/21}.
\end{equation}
In the latter (Coppi-like) case, which happens when
\begin{equation}\label{eqn:CoppiSmallTh}
\frac{a(t_\mrm{tr})}{a_0}\lesssim\biggl(\frac{m_e}{m_i}\biggr)^{1/2}\biggl(\frac{d_i}{a_0}\biggr)^{4/7}\betar^{-2/7}\Mach^{1/7},
\end{equation}
the maximum growth rate is
\begin{equation}\label{eqn:gammaCoppiSmallTh}
\gamma_\mrm{t,m}^\mrm{Coppi}\tauAr\sim\biggl(\frac{m_e}{m_i}\biggr)^{1/5}\betar^{-2/7}\Mach^{1/7}\Biggl(\frac{d_i^{4/7}a_0^{3/7}}{a}\Biggr) ,
\end{equation}
and so the critical time $t^\mrm{Coppi}_\mrm{cr}$ occurs when $\gamma_\mrm{t,m}^\mrm{Coppi}\taucs\sim{1}$, or
\begin{equation}\label{eqn:Coppidisrupts}
\frac{a(t^\mrm{Coppi}_\mrm{cr})}{a_0}\lesssim \biggl(\frac{m_e}{m_i}\biggr)^{1/10}\biggl(\frac{d_i}{a_0}\biggr)^{2/7}\betar^{-1/7}\Mach^{-3/7}.
\end{equation}
If the smallest parameter in the problem is $d_i/a_0$, so that (\ref{eqn:FKRdisrupts}) occurs before (\ref{eqn:CoppiSmallTh}) (i.e.~$t^\mrm{ FKR}_\mrm{cr} < t_\mrm{tr}$), then the CS will go unstable to mirror-stimulated FKR-like modes before the fastest-growing mode enters the large-$\Delta'$ regime. In this case, the critical CS thickness, $a_\mrm{cr}$, is given by \eqref{eqn:FKRdisrupts}. Comparing this to the expression for $a_\mrm{cr}$ when pressure anisotropy is {\em not} considered, equation \eqref{eqn:tearingDisrupts}, we see that mirrors increase $a_\mrm{cr}$ by a factor of ${\sim}(d_i/a_0)^{-2/7}\betar^{-4/7}\Mach^{2/21}$. If, instead,  $t^\mrm{FKR}_\mrm{cr} > t_{\rm tr}$, then the fastest-growing mirror-stimulated tearing mode becomes Coppi-like before tearing onsets, and $a_\mrm{cr}$ is effectively increased by a factor of ${\sim}(m_e/m_i)^{-1/15}(d_i/a_0)^{-8/21}\betar^{-1/7}\Mach^{-2/21}$.

%
%
\subsubsection{Mirror-stimulated tearing for $\theta\sim{x}_\mrm{m}/a$} 

If \eqref{eqn:smallThReq} is not satisfied, then the innermost mirror does not reach the center of the CS (i.e.~$k_{x,\mrm{m}}x_\mrm{m}\gg{1}$). Instead, the mirrors closest to the center with growth rate comparable to $\taucs^{-1}$ are most important, i.e.~those located at $x_\mrm{m}\sim\theta{a}$ (see \eqref{eqn:Deltaapprox}). Then the scaling laws in the previous section are modified; equations \eqref{eqn:gammaFKRsmallTh}--\eqref{eqn:Coppidisrupts} become, respectively,
\begin{gather}
\gamma_\mrm{t,m}^\mrm{FKR}\tauAr\sim\biggl(\frac{m_e}{m_i}\biggr)^{1/2}\biggl(\frac{d_i}{\theta a}\biggr)^2, \\
\label{eqn:FKRdisruptsTha}
\frac{a(t^\mrm{FKR}_\mrm{cr})}{a_0}\lesssim \biggl(\frac{m_e}{m_i}\biggr)^{1/6}\biggl(\frac{d_i}{\theta a_0}\biggr)^{2/3}\Mach^{-1/3}, \\
\label{eqn:CoppiTha}
\frac{a(t_\mrm{tr})}{a_0}\lesssim\biggl(\frac{m_e}{m_i}\biggr)^{1/2}\biggl(\frac{d_i}{\theta a_0}\biggr), \\
\gamma_\mrm{t,m}^\mrm{Coppi}\tauAr\sim\biggl(\frac{m_e}{m_i}\biggr)^{1/5}\biggl(\frac{d_i}{\theta a_0}\biggr), \\
\label{eqn:CoppiDisruptsTha}
\frac{a(t^\mrm{Coppi}_\mrm{cr})}{a_0}\lesssim\biggl(\frac{m_e}{m_i}\biggr)^{1/10}\biggl(\frac{d_i}{\theta a_0}\biggr)^{1/2}.
\end{gather}
Comparing \eqref{eqn:FKRdisruptsTha} and \eqref{eqn:CoppiTha}, we see that, if $d_i/(\theta a_0)\lesssim(m_e/m_i)^{-1}\Mach^{-1}$, tearing will onset before the fastest-growing mode can enter the large-$\Delta'$ regime (i.e.~$t^\mrm{FKR}_\mrm{cr}<t_\mrm{tr}$). In this case, $a_\mrm{cr}$ is given by \eqref{eqn:FKRdisruptsTha}, which is larger by a factor of ${\sim}\theta^{-2/3}$ than $a_\mrm{cr}$ derived without consideration of the mirrors, equation \eqref{eqn:tearingDisrupts}. Therefore, tearing will onset much sooner if $\theta\ll{1}$, whereas $t_\mrm{cr}$ is largely unaffected when $\theta\sim{1}$. If, instead, $t^\mrm{FKR}_\mrm{cr}>t_\mrm{tr}$, the mirror-stimulated tearing is Coppi-like, and $a_\mrm{cr}$ is given by \eqref{eqn:CoppiDisruptsTha}. However, the condition  \eqref{eqn:smallThReq} still must be satisfied, allowing only a narrow range of validity for $\theta$. Moreover, this range only exists if $d_i/a_0\gg(m_e/m_i)^{-7/4}\betar^{1/2}\Mach^{-2}$, a constraint not likely to be satisfied in the regime of interest. We therefore choose \eqref{eqn:FKRdisruptsTha} as the relevant condition for the onset of mirror-stimulated tearing for $\theta\sim x_\mrm{m}/a$.

%
%
\section{Discussion}\label{sec:summary}

While the specific quantitative model of CS evolution and mirror-stimulated tearing formulated herein is perhaps debatable, it nevertheless demonstrates an important, {\em qualitative} point: a gradually forming CS in a high-$\beta$, collisionless plasma easily produces enough pressure anisotropy to trigger the mirror instability, and the effect of this instability on the magnetic-field-line topology, and thus the tearing modes that instigate CS disruption via reconnection, ought to be considered.\footnote{In this respect, it is worth re-documenting the following prescient quote from the scarcely cited \citet{coppi83}: ``Thus we may consider the anisotropy-driven modes as a precursor of the spontaneous [tearing] ones and regard their effect as that of creating a region of macroscopic magnetic field turbulence near the neutral plane.''} For reasonable parameters, our theory predicts that the onset of reconnection in an evolving CS, driven by mirror-stimulated tearing modes, likely occurs earlier and at smaller scales than it would have without the mirrors, thereby placing a tighter upper limit on the aspect ratio of any forming CS (e.g.~compare \eqref{eqn:FKRdisrupts}, \eqref{eqn:Coppidisrupts}, and \eqref{eqn:FKRdisruptsTha} for the critical CS thickness at which mirror-stimulated tearing onsets to their $\Delta_p=0$ counterpart, equation \eqref{eqn:tearingDisrupts}). Whether or not these mirror-stimulated tearing modes ultimately grow to amplitudes $w\sim{a}_\mrm{eff}$, and perhaps beyond to ${\sim}a$ via island coalescence, to disrupt the CS awaits further work.

An immediate practical implication of this result is that numerical simulations of collisionless reconnection in high-$\beta$ plasmas should not initialize with a Maxwellian plasma embedded in an equilibrium CS. Instead, the CS should be allowed to evolve, and the particle distribution function self-consistently with it. A natural testing ground for this theory is the kinetic magnetorotational instability (MRI) \citep{quataert02,hq14}, thought to be the main driver of turbulence and enhanced transport in collisionless accretion flows, such as that onto the supermassive black hole at the Galactic center \citep{sharma06}. Historically, the linear MRI, at least in its MHD guise \citep{bh91}, was quickly shown to be a nonlinear ``channel'' solution in a differentially rotating disk \citep{gx94}, and various studies followed that employed Kelvin-Helmholtz and tearing ``parasitic'' modes to disrupt the otherwise resilient channels. In some theories, this disruption is credited for setting the steady-state level of magnetorotational turbulence as a function of the dissipative properties of the underlying magnetized fluid \citep[e.g.][]{pg09}. Given that the kinetic MRI both linearly and nonlinearly drives pressure anisotropy \citep{squire17}, it is worthwhile to contemplate a similar sequence of events, in which the kinetic MRI breaks down due to tearing modes stimulated by ion-Larmor-scale mirrors. Kinetic simulations of the MRI \citep[e.g.][]{riquelme12,hoshino13,hoshino15,kunz16,inchingolo18} may already be capable of testing this idea.

%
%
\begin{acknowledgments}
Support for A.A.~and M.W.K.~was provided by U.S.~DOE contract DE-AC02-09CH11466. This work benefited greatly from conversations with Nuno Loureiro, Alexander Schekochihin, and Dmitri Uzdensky, and from comments by the anonymous referees.
\end{acknowledgments}


\end{document}